\documentclass{article}
\usepackage{amsmath}
\usepackage{graphicx, geometry}
\usepackage{hyperref}

\begin{document}

\title{On extracting momentum from supports: \\
the bullet in the beam problem}
\author{Xiaoyu Zheng, Tianyi Guo and Peter Palffy-Muhoray \\
Department of Mathematical Sciences, Kent State University}
\maketitle

\begin{abstract}
To gain insights into momentum transfer from the supporting environment, we
consider the simple problem of a bullet, fired from below, into a wooden
beam resting on two supports. The resulting upward velocity of the beam
strongly depends on where the bullet enters the beam; this dependence is due
to upward momentum extracted by the beam from its supports. Our simple
example illustrates the momentum transfer mechanism exploited in the
remarkable dynamics of the chain fountain.
\end{abstract}

\section{Introduction}

The extraction of momentum from supporting surfaces is common enough to
escape notice, yet it deserves attention. We continuously extract 
momentum from the floor when we stand on it; fortunately, almost always this
extracted momentum cancels the momentum continuously injected into us by the
earth's gravitational field. We extract more when jumping up and while
landing; we extract horizontal momentum when accelerating and angular
momentum when turning. Strategies for extracting momentum, particularly by
inanimate objects, can be intriguing. The remarkable chain fountain is
lifted above the edge of its container by upward momentum acquired from the bottom of its
supporting container \cite{biggins2014understanding}; it is subsequently
pulled downward, in addition to gravity, via downward momentum acquired from the
floor \cite{corbin2018impact}. The falling chain gains energy from gravity,
and dissipates energy in inelastic collisions between the segments and between
the segments and supports.

\begin{figure}[h]
\centering 
\includegraphics[width=0.65\textwidth]{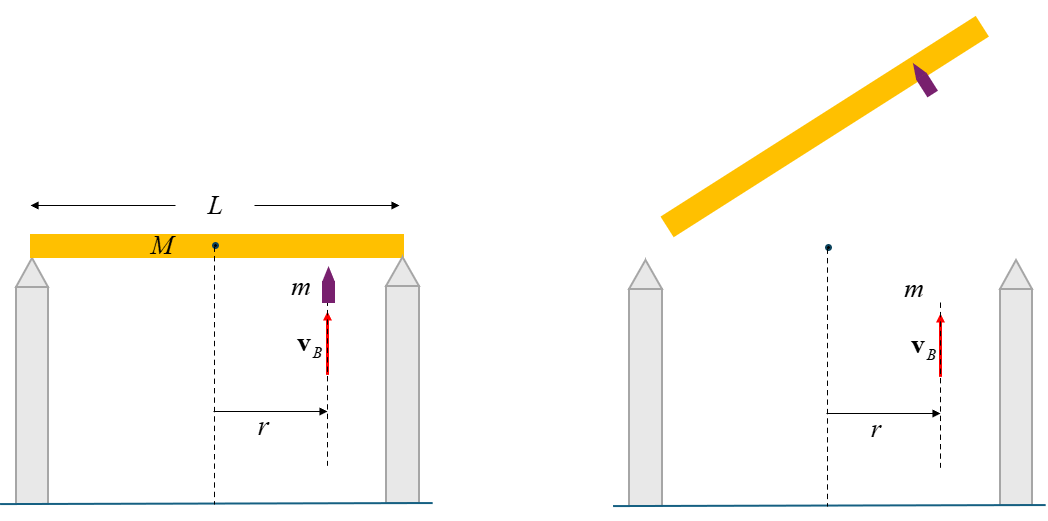} 
\caption{Schematic showing the arrangement of the wooden beam and the two
supporting fenceposts. Left image is before the bullet strikes, and the
right one is after.}
\label{fig:1}
\end{figure}

In this paper, we consider a simple experiment which illustrates the details of this
particular strategy for extracting momentum from a support. We consider a
bullet traveling upwards and striking a horizontal wooden beam supported at
its ends by two fenceposts, as shown in Fig.~\ref{fig:1}. On impact, the bullet becomes embedded in the wood; energy is dissipated in
this perfectly inelastic collision while momentum is transferred to the new
beam-bullet system. The question of interest is this: Where should the
bullet strike the beam in order for its center of mass (CoM) to rise as high
as possible? We solve this problem to understand the details of the momentum
transfer.

We remark that a variety of similar, but fundamentally different problems
are available in the literature. The physics of the well-known Veritasium Bullet–Block video, \url{https://www.youtube.com/watch?v=vWVZ6APXM4w}, in which a bullet is fired upward into a wooden block, has been examined and discussed in Ref.~\cite{mungan2015impulse}. In that setup, the block is supported only at its center by a horizontal axis which can freely ascend and about which the block can freely rotate without constraints. The bullet is fired into the block at some point along a line perpendicular to the axis, unlike the configuration considered here. In-depth studies of collisions and
impacts are available in the advanced texts of Goldsmith \cite%
{goldsmith2001impact} and Stronge \cite{stronge2018impact}, while
introductions to these subjects are given in \cite{semat1958physics} and
elsewhere, but we were unable to locate our example in the literature. A
classic problem, closely related to ours, is that of a bullet fired horizontally into a
beam on a frictionless surface; this problem is discussed by Kleppner and
Kolenkow \cite{kleppner2014introduction}, Halliday and Resnick \cite%
{halliday2013fundamentals} and Serway and Jewett \cite{serway2000physics},
but the ends of the beam are unconstrained. 

The unilateral constraint of the ends of the beam is the central feature of our problem. We carry out the analysis in Section 2 and follow with a discussion of the
results in Section 3.

\section{Analysis}

\subsection{Outline}

In the experiment, the bullet embeds itself in the horizontal beam supported by fenceposts, transferring both linear and angular momentum to
the beam-bullet system. The subsequent motion involves both translation and rotation, but
the contribution of the supporting fenceposts to the dynamics needs to be treated carefully.

The dynamics begins when the bullet becomes embedded in the beam, which we
assume occurs instantaneously \footnote{%
This is a major assumption. It takes a finite time for information from the
bullet to reach the supports, but likely this information gets there before the
embedding is complete.}. During this process, both linear and angular
momentum of the beam-bullet system are conserved, whereas the kinetic energy is not. Conservation of momenta determine the initial velocity of CoM and angular
velocity about the CoM immediately after impact, and thus fixes the post-impact kinetic energy of the beam-bullet system.

Following embedding, the beam begins to rise and rotate and may, depending on the impact location, collide almost immediately with a supporting fencepost. Such a collision allows the support to exert an impulse, so linear momentum is not conserved. However, because the fencepost is stationary and the contact is idealized as instantaneous and non-dissipative,  both mechanical energy and angular momentum about
the support are conserved, which together relate the translational and rotational motion before and after the collision.

\subsection{Before collision with the fencepost}

Changing the location of the point of impact of the bullet changes the
position of the CoM and this change significantly complicates the
description of the system dynamics. In order to have a simpler description
which still clearly shows key aspects of the problem, we assume that the
bullet mass $m$ is negligible compared to the mass $M$ of the beam, while
the linear momentum $p$ of the bullet is fixed. The CoM is then fixed at the
center of the beam, at a distance $L/2$ from the supports.

The bullet strikes the beam at distance $r$ from the CoM, as shown in Fig.~%
\ref{fig:1}. Prior to striking the beam, the bullet carries upward linear
momentum $p$. After lodging in the beam, linear momentum conservation gives
the initial upward velocity $v_{i}$ of the CoM, 
\begin{equation}
v_{i}=\frac{p}{M}.
\end{equation}%
The moment of inertia $I_{c}$ of the beam and bullet system about the CoM is 
\begin{equation}
I_{c}=\frac{1}{12}ML^{2}.
\end{equation}%
After the bullet is embedded, angular momentum conservation gives the
initial angular velocity $\omega _{ci}$ of the system about the CoM 
\begin{equation}
\omega _{ci}=\frac{rp}{I_{c}}=12\frac{rp}{ML^{2}}.
\end{equation}%
%

The kinetic energy of the beam-bullet system is%
\begin{eqnarray}
\mathcal{E}_{kin} =\frac{1}{2}Mv_{i}^{2}+\frac{1}{2}I_{c}\omega _{i}^{2} =\frac{1}{%
2}\frac{p^{2}}{M}(1+12\frac{r^{2}}{L^{2}}).
\end{eqnarray}%
The parabolic dependence on $r$ leads to a remarkable fourfold increase in
kinetic energy between $r=0$ to $r=L/2$. This increase of system energy is
due to the quadratic reduction of energy dissipation in the perfectly
inelastic beam-bullet collision with $r$.

The angular momentum of the system $\mathcal{L}_{f}$ about the tip of the
left fencepost is 
\begin{equation}
\mathcal{L}_{f}=I_{c}\omega _{ci}+\frac{1}{2}LMv_{i}=\frac{1}{2}Lp(1+2\frac{r%
}{L}).
\end{equation}%
It doubles in magnitude as $r$ goes from $r=0$ to $r=L/2$.

\subsection{The collision with the fencepost}

If the embedded bullet is to the right of center, as shown, the torque from
the bullet will be counterclockwise. As the bullet transfers its momentum to
the beam, the CoM of the beam begins to rise and the beam begins to rotate.
The upward speed of the left end of the beam is%
\begin{equation}
v_{tip}=v_{i}-\frac{L}{2}\omega _{ci}=\frac{p}{M}(1-6\frac{r}{L}).
\end{equation}%
If $r<L/6$, rotation is slow and $v_{tip}>0$, and the beam rises and
rotates. It loses contact with the fencepost immediately, and does not
regains contact again while ascending. If $r>L/6$, however, rotation is
faster and $v_{tip}<0$. This indicates that fencepost represents a
constraint, and that the solution $v_{i}$ and $\omega _{ci}$ are non longer
physical. Practically, it indicates that the left end of the beam
collides with the left fencepost.

In either case, the two conserved quantities of the beam-bullet system
in the subsequent motion are kinetic energy $%
\mathcal{E}$ and the angular momentum of the system $\mathcal{L}_{f}$ about
the tip of the left fencepost.

\subsection{After collision with the fencepost}

Upon collision of the beam with the fencepost, both the linear velocity
and the angular velocity of the CoM change. Denoting the unknown new values
of these with the subscript $f$, the conservation of kinetic energy and angular
momentum about the left fencepost give%
\begin{equation}
\frac{1}{2}Mv_{i}^{2}+\frac{1}{2}I_{c}\omega _{ci}^{2}=\frac{1}{2}Mv_{f}^{2}+%
\frac{1}{2}I_{c}\omega _{cf}^{2},
\end{equation}%
and 
\begin{equation}
I_{c}\omega _{ci}+\frac{1}{2}LMv_{i}=I_{c}\omega _{cf}+\frac{1}{2}LMv_{f},
\end{equation}%
which can be solved simultaneously.

One set of solutions is 
\begin{eqnarray}
v_{f} &=&v_{i}=\dfrac{p}{M}, \\
\omega _{cf} &=&\omega _{i}=\dfrac{3p}{ML}(\frac{4r}{L}),
\end{eqnarray}
when there is no collision or there are no supports, and the motion
continues.

There is an alternate set of solution: 
\begin{eqnarray}
v_{f} &=&\frac{1}{4}L\omega _{i}+\frac{1}{2}v_{i}=\frac{p}{M}(3\frac{r}{L}+%
\frac{1}{2}), \\
\omega _{cf} &=&-\frac{1}{2}\omega _{i}+3\frac{1}{L}v_{i}=\frac{3p}{ML}(1-2%
\frac{r}{L}).
\end{eqnarray}%
This set of solutions describes the dynamics after collision of the beam
with the left fencepost. It is straightforward to show that the collision of
the beam end with the left fencepost is always perfectly elastic\cite%
{palffy2019paradox}, consistent with our assumption of energy conservation.

\begin{figure}[h]
\centering 
\includegraphics[width=0.95\textwidth]{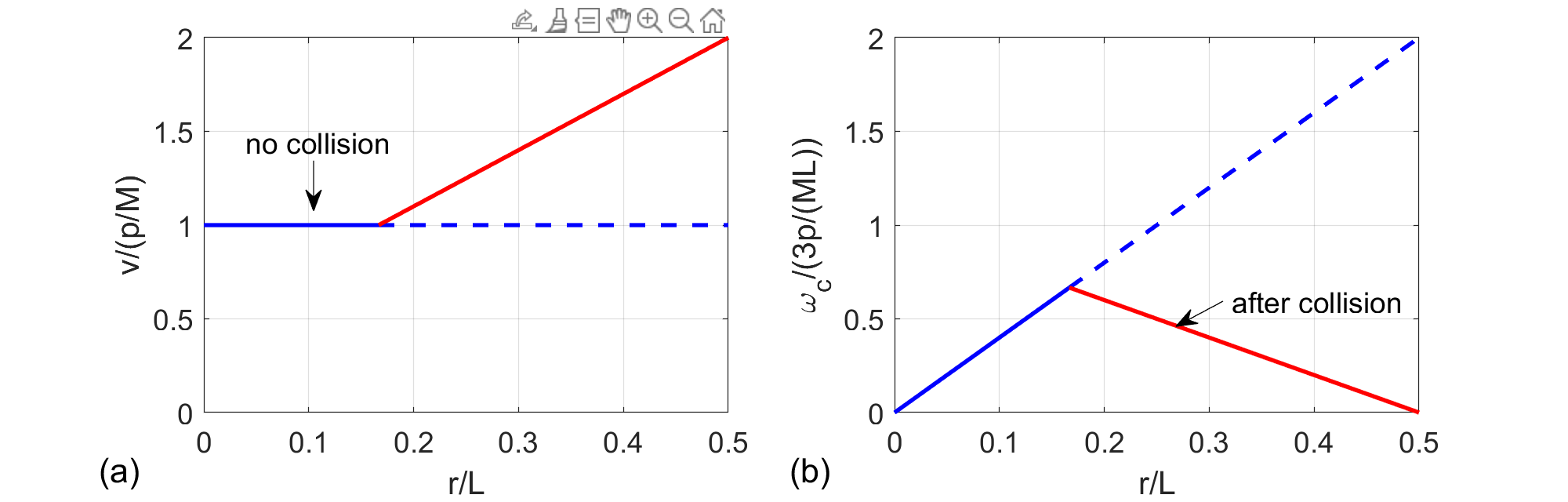} 
\caption{(a) The CoM velocity $v$ vs.~$r$. (b) The
angular velocity $\protect\omega_c$ vs.~$r$.}
\label{fig:v}
\end{figure}

Fig.~\ref{fig:v} shows the CoM velocity $v$ and the angular velocity about
CoM $\omega _{c}$ in terms of $r$. For $r<L/6$, the system is on the
\textquotedblleft no collision" branch, where the CoM rises with constant
speed $v_{i}$, regardless of $r$. The beam rotates counterclockwise with
angular velocity $\omega _{i}$, and the speed of rotation increases linearly
with $r$. The two solution branches cross when $r=L/6$, this is the point
where the tip is at rest on the left fencepost as the CoM rises. For $r>L/6$%
, the system is on the \textquotedblleft after collision" branch. The
fencepost injects upward momentum into the beam. The CoM rises with speed
which increases linearly with $r$, reaching the remarkable value of $2p/M$,
when $r=L/2$, which is twice that of $r=0$. The speed of rotation on this
branch decreases linearly with $r$, reaching zero at $r=L/2$. At $r=L/2$,
the momentum acquired from the fencepost equals the initial momentum of the
bullet. 

\section{Discussion}

We note again that the bullet always transfers the same upward momentum to
the beam, regardless of where it strikes. If it strikes in the middle, the
beam will simply rise, but not rotate. As the point where the bullet strikes
moves away from the center towards the right, the beam begins to rotate; the
upward velocity remains constant but angular velocity increases until the
distance from center is one-sixth of the total beam length. Beyond this
point, the upward velocity increases while the angular velocity decreases
linearly with distance due to collision with the left fencepost. When the
bullet strikes the beam at its right end, there is no rotation, and the beam
rises twice as fast than when the bullet strikes in the center.

The kinetic energy of the system increases with distance of the bullet from
the center. The source of the increasing energy is the parabolic reduction
of energy dissipation in the beam-bullet collision with distance. When the
distance is less than one-sixth of the beam length, this increased energy
goes into rotation of the beam. Beyond this distance, the rotation is
reduced, and the increased energy increases the upward velocity.

From the momentum perspective, the left fencepost transmits the linear
momentum to the beam-bullet system so that it raises higher after the
collision. From the energy perspective, the support enables the conversion
from rotational kinetic energy to translational kinetic energy.

In our analysis, gravity did not play a role. The height to which the CoM would rise in the
presence of gravity is given by equating the kinetic energy of translation
here with the gravitational potential energy. 

If the beam was supported not at the ends, but at two intermediate points,
then multiple collisions with supports, as well as horizontal translation,
could take place. Multiple collision scenarios are interesting indeed; for
the sake of brevity, they will not be considered here.

\begin{figure}[h]
\centering 
\includegraphics[width=0.8\textwidth]{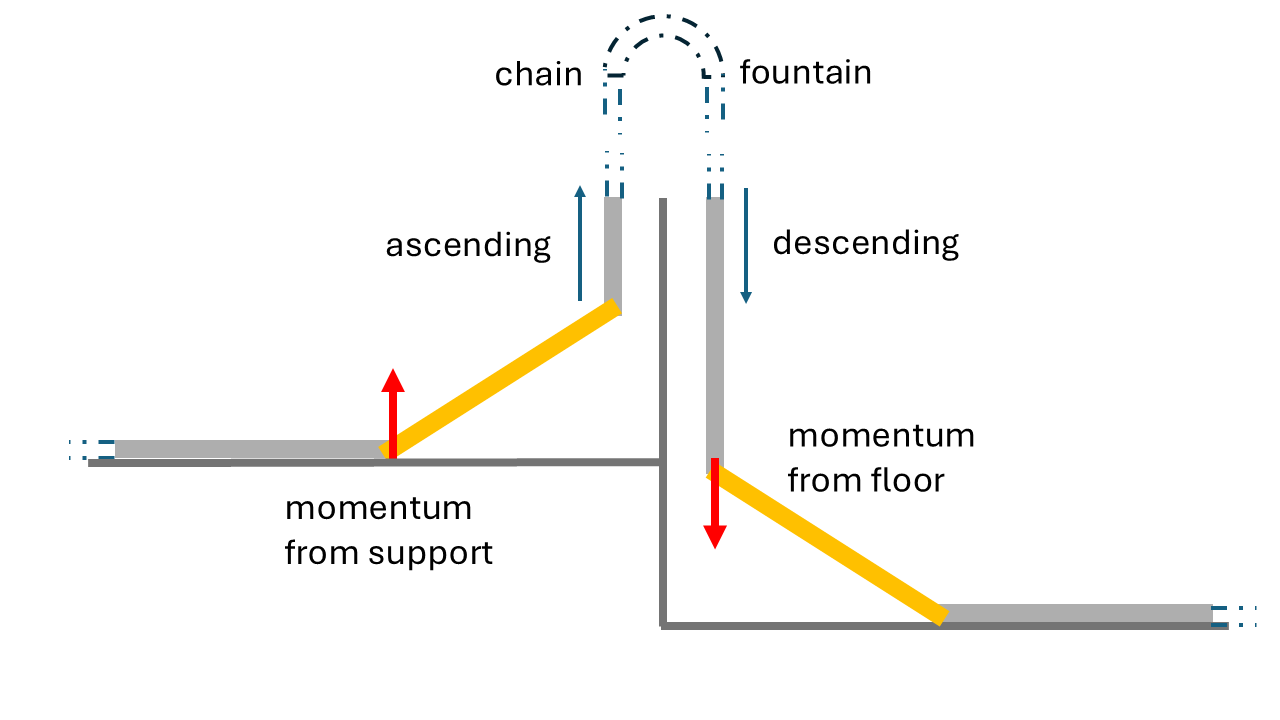} 
\caption{Schematic of the parts of chain fountain. Upward momentum is
transferred from the support to the chain on the left, and downward momentum is transferred from the floor to the chain on the right.}
\label{fig:fountain}
\end{figure}

We note that chains used in chain fountain and similar experiments can be
regarded as consisting of jointed rigid segments, such as our beam. Lifting
one end of such a rigid segment on the ascending side is equivalent to the
impact of a bullet, and the resting end extracts upward momentum from the
supporting surface. In viewing the descending side from a freely falling
inertial frame, the floor hitting the end of the segment is equivalent to the
impact of a bullet, and the upper end of the segment injects downward momentum, originating in the floor,
into the segment above it.  Of course, there are many other details to consider
as was thoroughly done in \cite{biggins2014understanding}, but the essential
momentum transfer aspects of the chain fountain are the same as those in our
simple problem.

This simple experiment gives insight into strategies whereby simple systems
extract momentum from their environment. Our example shows that, by shooting
the bullet into one end of the beam rather than into the center, the upward
CoM velocity increases by a factor of four, due to momentum extracted from
its environment. Essentially the same strategy is utilized by the chain
fountain in extracting upward momentum from its container\cite%
{zheng2018acquiring}. We hope that the example discussed here may be of use
in considering other momentum transfer processes.

%

\section*{Conflict of Interest}

The authors have no conflicts to disclose.

\bibliographystyle{plain}
\bibliography{New2}

\end{document}